\documentclass[rms3]{article}
 
\usepackage{graphics}
\usepackage{rms3}

\begin{document}

\rms{V. SCHROEDER {\it et al.}}
 	{CORONA EMISSION FROM HYDROMETEORS}
	{1998}
	{X}
 	{1}
	{X}

\title{A model study of corona emission from hydrometeors}
 
\author{Vicki Schroeder$^1$\footnote{Corresponding author: Geophysics Program,
University of Washington, Box 351650, Seattle, 98195-1650, USA email:vicki@geophys.washington.edu}, M.B. Baker$^1$ and John Latham$^2$}
 
\affiliation{$^1$University of Washington, USA\\               
    $^2$National Center for Atmospheric Research, USA}
 
\date{(submitted for review)}
 
\summary{
The maximum measured electric fields in thunderclouds are an order of magnitude less than
the fields required for electric breakdown of the air. One explanation for lightning
initiation in these low fields is that electric breakdown first occurs at the surfaces of raindrops
where the ambient field is enhanced very locally due to the drop geometry . Laboratory
experiments [Crabb \& Latham, 1974] indicate that colliding raindrops which coalesce to form
elongated water filaments can produce positive corona in ambient fields close to those
measured in thunderclouds.
 We calculate the E-field distribution around a simulated coalesced drop pair and use a
numerical model to study the positive corona mechanisms in detail.
Our results give good agreement with the laboratory observations.
At the altitudes (and thus low pressures) at which lightning initiation is
observed, our results show that positive corona can occur at
observed in-cloud E-fields.} 
 
 %\keywords{Keyword one\ksp Keyword two\ksp Keyword three}

\ahead{Introduction}
\label{intro}

Lightning initiation in thunderclouds is poorly understood. An order
of magnitude discrepancy exists between the maximum measured electric fields
(E-fields) in clouds
($E_{max}$) and the E-fields required for dielectric breakdown of air.
$E_{max} \sim$ 100 - 400 kV/m  \cite{marshall, winn}. $E_{breakdown} \sim$ 2700
kV/m at surface pressure;  at the lower pressures ($\sim$ 500 mb)  at which most
lightning is observed to initiate, $E_{breakdown}$ is reduced to $\sim$ 1400
kV/m - still far greater than $E_{max}$.   

One of several explanations put forward to explain this discrepancy is the
enhancement of the electric field near the surfaces of hydrometeors (water or
ice particles) in clouds. A set of laboratory experiments by Crabb \& Latham 
(hereafter CL) showed that colliding raindrops may provide the starting point for lightning
initiation.  CL obtained very promising results in a set of experiments in which they
measured the E-fields required to initiate a discharge from the
surface of filamentary, temporarily coalesced drops created when two water drops
collided. They observed pulsed, intermittent discharges in a localized region
near surface of the drop and found that the E-fields required lay within the
range of observed thunderstorm E-fields. 

We extended models of the positive discharge process developed by
\cite{dw65, gal79, bondiou, abdel-salam} in order to study the discharge
processes occurring from hydrometeors. CL's laboratory conditions were used to
initialize the model and their results were used to validate it.  
With our model we were able to vary both the microphysical and environmental
conditions and investigate a range of conditions applicable to those found in
thunderclouds. In particular, we investigated
continuous discharges from the drop surface and we studied  the pressure
dependence of discharge initiation E-fields.  

We begin with a brief description of Crabb \& Latham's experimental procedures and
results, followed by a discussion of the discharge model used to make the
 calculations. Finally we discuss our model results - showing the E-field required to
 initiate various discharge types as a function of the coalesced drop properties
and air pressure.

\ahead{Laboratory Experiments}
\label{lab}

\begin{figure}[ht]
\begin{center}
\includegraphics{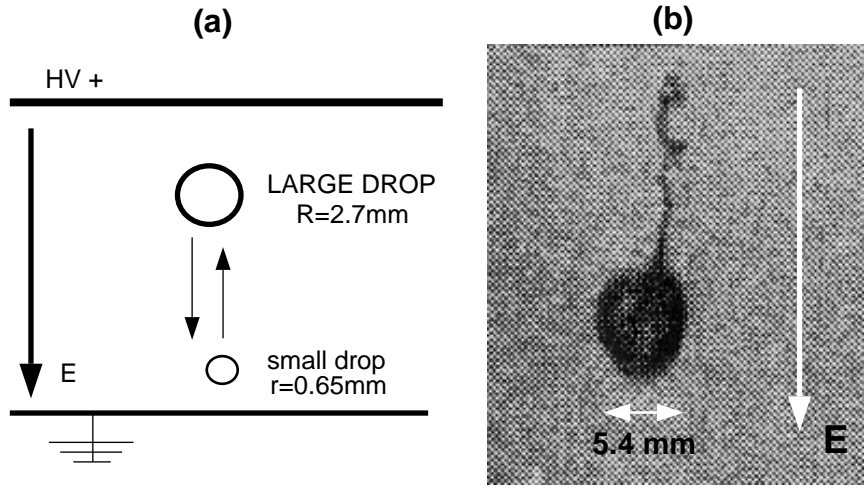}
\end{center}
\caption{ (a) Schematic of Crabb \& Latham's experimental setup in which two
drops (R=2.7mm and r=0.65mm) collided in the presence of an applied electric
field. (b) Photograph of the coalesced drop that formed after the collision.}
\label{clsetup}
\end{figure}

Figure \ref{clsetup} (a) shows a schematic of the CL
experiment . Their chamber, held at surface pressure, had a
positive, high voltage upper plate and a grounded lower plate separated by
50mm. Voltages of up to 30 kV could be applied - corresponding to a maximum
uniform E-field of 600 kV/m within the chamber. Large water drops (R=2.7mm) were
dropped into the chamber and collided with small drops (r=0.65mm) which were
blown upwards, simulating drops moving in updrafts in
thunderclouds. A variety of coalesced drop shapes were observed, depending on the
nature of the collision. CL described three basic collision modes: head-on,
glancing and intermediate. Glancing and intermediate collisions produced a
coalesced drop with a long filament extending from the 
large drop - see Fig \ref{clsetup} (b).   Head-on collisions
resulted in a flattening of the large drop and did not produce these long
filaments.  The drops remained in the coalesced state for $\sim$ 1 ms.   

In CL's setup, a negative charge was induced on the upper surface of the drop while the lower
end had a positive induced charge. In the thundercloud setting these drops would
be located above the negative charge center of the cloud. CL recorded the size and shape of the
coalesced drops as well as the applied E-fields required to initiate discharges for a
large number of coalesced drops. They observed discharges from both ends
of the drop but focussed on the positive pulses occurring at the lower surface
of the drop. This surface was observed to remain intact. In contrast surface disruption was
observed at the upper, negative surface of the drop.

CL observed that positive burst pulses occured for values of E between 250  and 500
kV/m, depending on the length of the coalesced drop. 

\ahead{Model Description}
\label{model}

We describe three basic discharge processes that can occur at the
surfaces of drops in the presence of strong E-fields. 

The first process, {\em surface disruption} discharge, occurs when the
electrostatic repulsive force on a drop in a strong E-field  exceeds the surface
tension. This results in breakup of the drop surface, and an associated
discharge. \cite{daw69} observed the surface E-field, 
$E_{disruption}$, required to initiate this form of discharge as a function of
drop size.  $E_{disruption}$ is independent of pressure. 

The remaining processes are referred to as 'pure' corona processes because the
discharge initiates without the occurrence of drop surface disruption.  
These processes are:
\begin{itemize}
\item {\em burst pulse} discharges, which are intermittent, and
\item {\em continuous streamers}, which are capable of propagating continuously.
\end{itemize}
We discuss each in detail below. All of these processes result in
deposition of charge on the drop: either positive or negative charge depending on
the sign of $E_{external}$.   We focussed only on positive corona:
it is simpler to model, has a lower initiation threshold and was studied more
extensively by CL than negative corona. 

\bhead{Positive Pure Corona model}
\label{purecorona}

\begin{figure}[ht]
\begin{center}
\includegraphics{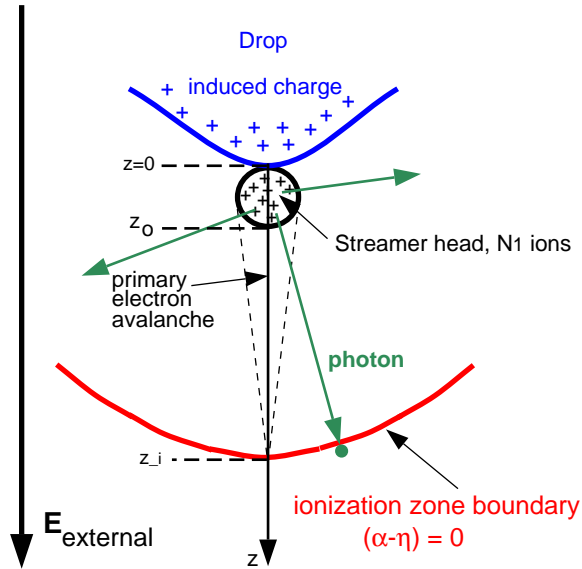}
\end{center}
\caption{Schematic of positive discharge formation near the
surface of a drop.  Free electrons are accelerated by the E-field and undergo collisions
with air molecules. The ionization of molecules within the ionization zone leads
to an exponential growth of electrons (avalanche)  and the formation of a
spherical streamer head.} 
\label{streamer_cartoon}
\end{figure}

Following \cite{dw65, gal79, bondiou} we model the positive discharge as a
series of electron avalanches. 

Consider the electric field near the surface of a drop which is 
situated in an external electric field, $E_{external}$. (See Fig
\ref{streamer_cartoon}.)
The total electric field is a function of the distance, $r$, from the drop
surface.  Initially, the total electric field at $r$ is
\begin{eqnarray}
E (r) = E_g(r) = {E_{external} + E_{drop} (r)}
\label{Eg}
\end{eqnarray} 
where  $E_{drop}$ is the contribution due to charge induced on the drop. $E_g (r)$ is called
the geometric field.  

In the presence of $E$ free electrons are accelerated and undergo collisions
with air molecules.  At some radial distance from the drop $E$ is such that:
\begin{equation}
\label{boundary}
 \alpha(E/p) =  \eta(E/p)
\end{equation}
where $\alpha$ [m$^{-1}$] and $\eta$ [m$^{-1}$] are the ionization and
attachment coefficient for electrons in air, respectively and p is the total air pressure. 
The surface defined by eqn (\ref{boundary}) is the {\em ionization zone boundary} -
inside this boundary $\alpha > \eta$ and there is a net growth of free
electrons.  At surface pressure the  ionization zone boundary is the
surface along which $E = 2700$ kV/m. Figure \ref{alphaeta} shows $\alpha$ and
$\eta$ as functions of $E/p$  \cite{geballe,loeb, badaloni, ibrahim}.  

\begin{figure}[btp]
\begin{center}
\includegraphics{aefm.epsi}
\end{center}
\caption{(a) Ratio of ionization coefficient to pressure, $\alpha/p$,  for electrons in air
\cite{badaloni, loeb}
(b) Ratio of attachment  coefficient to pressure, $\eta/p$, for electrons in air \cite{badaloni, geballe}
(c) $\Psi = f_1 \cdot f_2 \cdot \theta$ where $f_1$ is the number of photons
created per ionizing collision, $f_2$ [m$^{-1}$] is the number of photoions
created per photon per meter and $\theta$ is a solid angle = 2$\pi$ in our
calculations \cite{penney} 
(d) Ratio of photon absorption coefficient to pressure, $\mu/p$, in air \cite{penney}.}
\label{alphaeta}
\end{figure}

Following \cite{dw65,griffiths,gal79} we replace the three-dimensional problem by a
one-dimensional one in which all avalanches occur along the z-axis. 
The point $z_i$ marks the intersection of the ionization boundary with the $z$-axis.
When a free electron, starting at $z_i$, is accelerated by $E$ towards the drop,
the number of electrons grows exponentially as $z$ 
decreases. This is referred to as the {\em primary electron avalanche}. Due to the
exponential nature of the growth, most of the ionizing collisions occur near
the surface of the drop. The free electrons are then absorbed by the
drop, leaving behind a concentration of positive ions, modeled as a sphere
\cite{dw65, gal79}, and referred to as the {\em streamer head}.  

The number of positive ions formed by the primary avalanche traveling from the
ionization zone boundary, $z_i$, to the drop surface, $z_o$, is given by:
\begin{eqnarray}
{N_1} = exp[\int_{z_i}^{o} 
(\alpha (z)) - \eta(z)) dz]
\label{Ntotal}
\end{eqnarray}
The  radius of the streamer head is approximately:
\begin{eqnarray}
R_s = [6  \int \limits_{z_i}^{o} (D(z)/{v(z)}dz)]^{1/2}
\label{radius}
\end{eqnarray}
where $D$ and $v$ are the electron diffusivity and drift velocity,
respectively. $D$ and $v$ are functions of the ratio $(E/p)$ and thus depend on
$z$ \cite{healey, ibrahim}. $R_s \approx
30 \mu m$ at surface pressure for most of the calculations reported here.

The total electric field at $z$ is now given by:
\begin{eqnarray}
E (z) = {E_g (z) + {e N_1 \over {4 \pi \epsilon_0 ( z - R_s)^2}}}
\label{Etotal}
\end{eqnarray}
where the second term, $E_c$,  is the E-field due to the spherical charge concentration of
the streamer head. 

In addition to ionization, collisions between the free electrons and air molecules
also result in the excitation of the molecules, which then emit photons on
decay. A certain fraction of these photons in turn have sufficient energy to
ionize molecules that they encounter, creating {\em photoelectrons}. These
photoelectrons then start a series of {\em secondary avalanches} which converge
on the drop from all directions. 

The number of photoelectrons created per m at a radial distance, $l$, from the
drop surface is given by:
\begin{eqnarray}
P(l) = f_1 N_1 \cdot exp[-\mu l] \cdot f_2 \cdot G
\label{photoelectrons}
\end{eqnarray}
\begin{tabbing}
where 	\= $f_1$ is the number of photons created per ionizing collision\\
	\> $\mu$ [m$^{-1}$] is the photon absorption coefficient in air \\
	\> $f_2$ [m$^{-1}$] is the number of photoions created per photon per meter\\
	\> $G$ is a geometric factor to account for the fact that some\\
	\> \hspace{5mm} photons are absorbed by the drop.
\end{tabbing}
Both $\mu$ and $f_1 \cdot f_2$ are functions of $l \cdot p$, the product of the distance from
the photon source (the collisions) and air pressure \cite{penney} - see Fig \ref{alphaeta}.

Then the total number of ions created in the secondary avalanches is given by:
\begin{eqnarray}
N_2 = \int_{z_i}^{z_o} P(l) \cdot
exp[\int_{l}^{z_o} (\alpha - \eta) \, dz] \, \, dl
\label{secondaryions}
\end{eqnarray}
where $z_o$ indicates the position of the primary streamer head surface.

\chead{(i)} {Initiation Conditions}
\label{initiation}

A {\em burst pulse discharge} is initiated if the number of photoelectrons
created along the ionization zone boundary during the growth of the primary
avalanche is equivalent to the number of photoelectrons that started the primary
avalanche (commonly taken as 1) \cite{abdel-salam}. 

We consider photoelectron production in a region of depth (1/$\mu$) along the
ionization zone boundary and write the above condition as follows: 
\begin{equation}
\label{photocondition}
\frac{P(z_i)}{\mu(z_i)} = 1
\end{equation}
This type of discharge is intermittent because the number of
positive ions, $N_1$, in the primary streamer head is too small to attract the
following avalanches to its surface. Instead, the successor avalanches are directed to the
drop - allowing the discharge to ``spread'' over the drop surface.

The fulfillment of eqn (\ref{photocondition}) is strongly influenced by the relationship
between the mean free photon path (1/$\mu$) in air and the location of
$z_i$. For drops with small radii,  $z_i$ is closer to the drop
surface than for those with larger radii. Thus for small drops the number of photons likely to reach
the ionization zone boundary is increased.  The chances of a photoelectron being
produced then increases and it is easier to initiate a burst pulse discharge
under these conditions.  

A more stringent initiation condition exists for {\em continuous streamers}.  In
this case the number of positive ions in the primary streamer head must be large
enough to attract the secondary avalanches to the streamer head surface. This is achieved
when the radial E-field around the streamer head, $E_c  = \frac{e N_1}{4 \pi
\epsilon_0 ( z - R_s)^2} \sim E_g$ \cite{abdel-salam}. In addition:\\
(a) $N_2$, the number of positive ions in the streamer head that
results from the secondary avalanches, must equal $N_1$, the number of positive ions
created by the primary avalanche, and\\
(b) the radius of the secondary streamer head must equal $R_s$, the radius of the
primary streamer head. \\
These conditions ensure that the initial streamer head charge density is
reproduced in the second streamer head. Continued
reproduction of the streamer head in subsequent steps results in propagation of the positive
streamer away from the drop surface. 
For all the geometric conditions considered in this paper, the initiation of
continuous streamers requires a larger external E-field than for burst pulse
discharge initiation.

The minimum value of $E_{external}$ necessary to initiate a discharge at
pressure p is referred to as $E_{initiation}(p)$ and depends on the type of
discharge (burst pulse or continuous streamer).

\bhead{Model Procedure}

\begin{figure}[h]
\begin{center}
\includegraphics{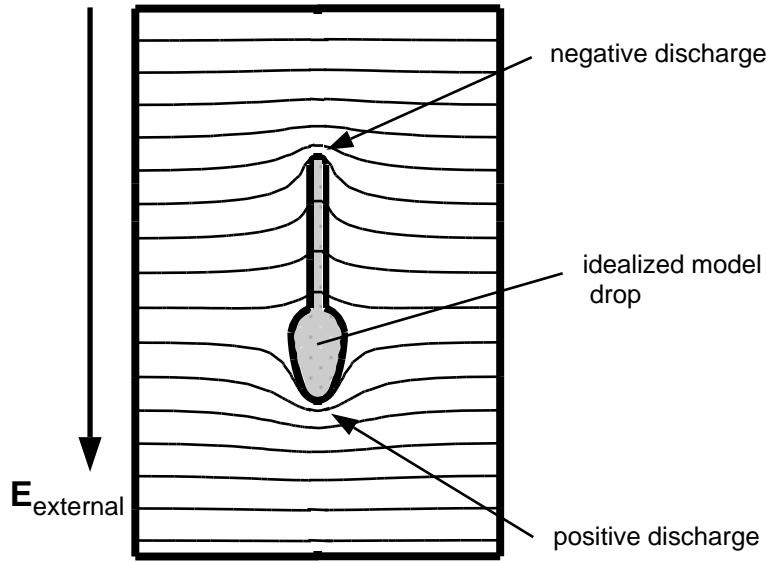}
\end{center}
\caption{Idealized, model drop used in this study with lines of equipotential
shown. The field distribution around the drop
is calculated using a finite element method \cite{quick}. }
\label{show_model}
\end{figure}

Our results were obtained using the following procedure:
\begin{itemize}
\item Define the drop shape and permittivity, $\epsilon$.   
The idealized shape used is shown in Fig \ref{show_model}.
Set the air pressure, p.
\item Apply $E_{external}$ to the drop.
\item Calculate the E-field distribution around the drop using a finite element
method based solving routine \cite{quick}. 
\item Compare the E-field at the drop's negative surface to the known surface
disruption field threshold, $E_{disruption}$ \cite{daw69}.\\
If $E(surface) > E_{disruption}$ then add varying amounts of positive
charge, $Q_{drop}$, to the drop. 
\item Recalculate the E-field distribution \cite{quick}.
\item Find the ionization zone boundary, $z_i$.
\item Compute $N_1$ and $R_1$ from eqns (\ref{Ntotal}) and (\ref{radius})
respectively.
\item  Compute $P(l)$ at $z_i$ from eqn (\ref{photoelectrons}).\\
For $\frac{P(z_i)}{\mu(z_i)} = 1, E_{external} = E_{initiation}(p)$ for burst pulse discharges.
\item  Compute $N_2$ and $R_2$ from eqns (\ref{secondaryions}) and (\ref{radius})
respectively.\\
If $E_c \sim E_g, N_2 = N_1$ and $R_2 = R_1$, then $E_{external} = E_{intiation}(p)$ for
continuous streamers.
\end{itemize}

\vspace{1cm}

\ahead{Results}
\label{results}

\bhead{Surface disruption}
For $E_{external} \ge$ 200 kV/m the calculated E-field at the surface 
of the upper, negative end of the drop was $\ge$ 8500 kV/m, the value of
$E_{disruption}$ at p=1000mb for a water drop of radius r =0.65mm \cite{daw69}. 
This is consistent with CL's observations that the upper surface disrupted
at these E-field strengths.  The resulting negative discharge then
deposited positive charge on the drop. 

\bhead{E$_{initiation}$ vs Q$_{drop}$}

\begin{figure}[hb]
\begin{center}
\includegraphics{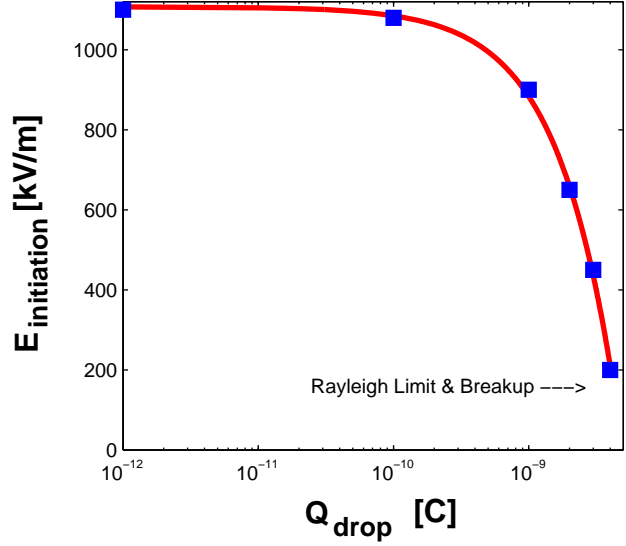}
\end{center}
\caption{$E_{initiation}$ for positive burst pulse discharges from the lower
positive end of the drop as a function of $Q_{drop}$, the charge deposited on
the drop by the negative corona from the upper end. The drop length is held fixed
at L=20mm.} 
\label{EvsQ}
\end{figure}

Fig \ref{EvsQ} shows the  $E_{initiation}$ values for positive burst pulse
discharges from the lower positive end of the drop as a function of the charge,
$Q_{drop}$, deposited on the drop by the negative discharge from the upper end. The
drop length is held fixed at L=20mm. 

 $E_{initiation}$ decreases rapidly once $Q_{drop}$ exceeds $10^{-10}$
C. The Rayleigh stability criterion \cite{rayleigh, taylor} gives $Q_{RL}$, the maximum
charge that a sphere of liquid can hold before the electrostatic repulsive force
overcomes the surface tension. In SI units it is given by:
\begin{equation}
\label{rayleigh}
Q_{RL}^2 = 64  \, \pi^2 \, \epsilon_o \cdot r^3 \, \sigma
\end{equation}
where  r is the sphere radius and $\sigma$ is the surface tension.

For our drop dimensions $Q_{RL} \approx 4 \times 10^{-9}$ C. Since CL  did not
observe disruption of the lower surface of the drop, we limited 
our calculations to $Q_{drop} < Q_{RL}$.  For larger allowed values of $Q_{drop}$, close
to the Rayleigh limit $Q_{RL}$, the values of $E_{initiation}$ become comparable to CL's
experimental values and to those observed in thunderclouds.

In addition to the burst pulse discharges we also calculated the fields required to
initiate continuous streamers.  For $Q_{drop}$ just below the
Rayleigh limit, $E_{initiation} \approx$ 400 kV/m for continuous streamers, approximately 50\% greater 
than that required for burst pulse discharges.

\begin{figure}[ht]
\begin{center}
\includegraphics{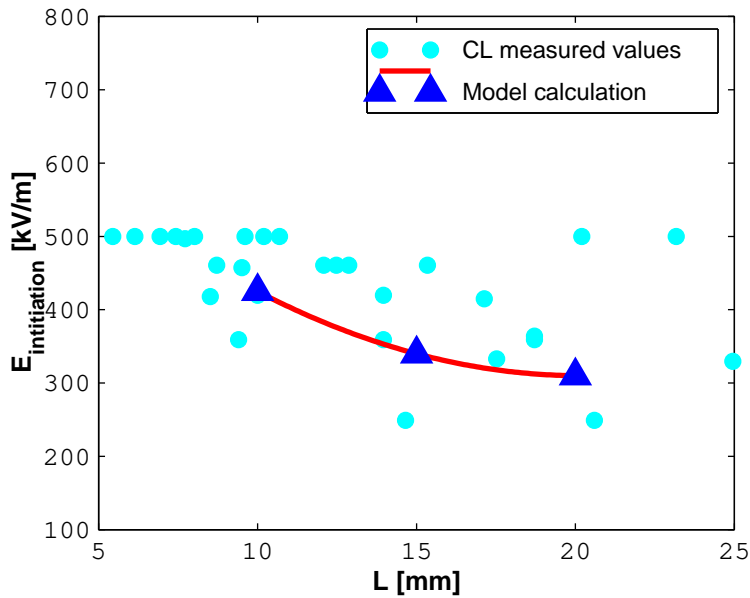}
\end{center}
\caption{$E_{initiation}$ for burst pulse discharge as a function of the drop
length L, for fixed charge density. Triangles: calculated values of $E_{initiation}$ for burst pulse
discharges. Circles: Crabb \& Latham's measured values.}
\label{EvsL}
\end{figure}

\bhead{E$_{initiation}$ vs drop length, L}

We now held the charge density, $\rho$, on the drop fixed at 0.035 C/m$^3$ and
varied the drop length, L. The circles in Fig \ref{EvsL} represent CL's
measured values. We found that our modeled values of $E_{initiation}$ for the
burst pulse discharges ($\bigtriangleup$) decreased with increasing L,
consistent with the trend that CL observed.  

The agreement between the calculated results and observation is promising and offers
validation of our model processes. The scatter in
CL's results is most likely due to either the differences in the shape of the lower
end of the coalesced drops or the amount of charge that is deposited by the
negative discharge.  We found higher $E_{initiation}$ values for a 
coalesced drop with a spherically shaped lower end, while lower
$E_{initiation}$ values were recorded for more pointed lower ends. Our
idealized shape with $\rho$=0.035 C/m$^3$, however, provided good
agreement with CL's average values for $E_{initiation}$.

\begin{figure}[ht]
\begin{center}
\includegraphics{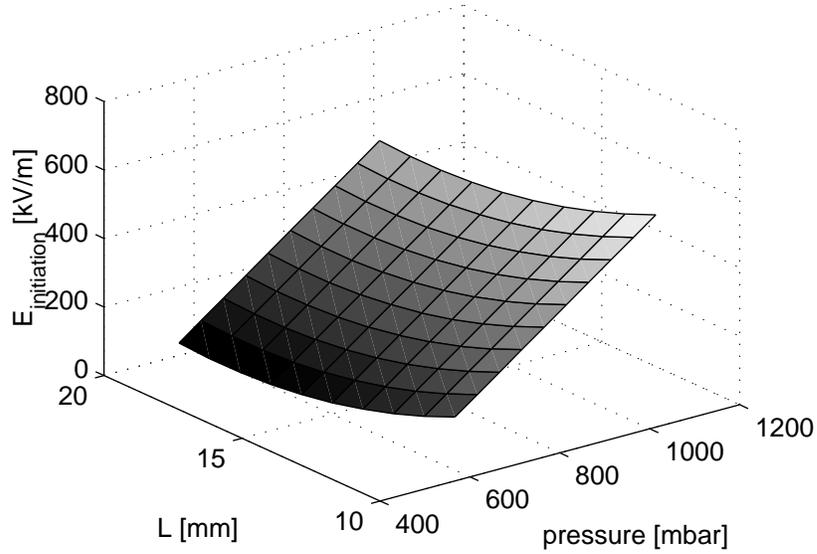}
\end{center}
\caption{$E_{initiation}$ for continuous discharges as a function of both
drop length, L [mm], and pressure, p [mb]}
\label{E3d}
\end{figure}

The same calculations were carried out for continuous
streamers and the results are shown in Fig \ref{E3d}. $E_{initiation}$ decreased
with L in much the same way as for burst pulses. The $E_{initiation}$ values for
the continuous streamers were, however, $\sim 50 \%$ larger than those required for burst pulses. 

\begin{figure}[h]
\begin{center}
\includegraphics{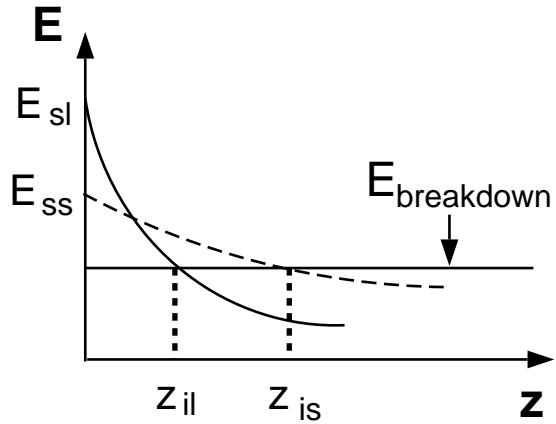}
\end{center}
\caption{E-field as a function of distance, z, from the surface of the
drop. E$_{sl}$ and E$_{ss}$ are the surface fields for long and short
drops, respectively. The ionization zone boundaries for long and
short drops are indicated by $z_{il}$ and $z_{is}$, respectively.}
\label{Evsz}
\end{figure}

Figure \ref{Evsz} shows the two competing processes that determine the
dependence of $E_{initiation}$ on L for continuous streamers.  On the one hand, for a given ambient
E-field, the surface field at the tip of the filament increases
with increasing L, which lowers $E_{initiation}$. In opposition to this, as L increases $E_g(z)$
decreases more rapidly with $z$, the distance from the surface. This
reduces the size of the ionization zone and thus increases
$E_{initiation}$. Fig \ref{E3d} shows that the former process
dominates; i.e. that the increased average field within the ionization zone compensates for the
electron's shortened path - leading to a lowering of $E_{initiation}$ as the
filament length is increased. As Fig \ref{E3d} 
indicates, $dE_{initiation}/dL$ decreases as L increases and the effect of increased
length becomes less significant for L$>$20mm.

\vspace{5mm}

\bhead{The pressure effect}

All CL's measurements were made at surface pressure (1000 mb). It is, however, of
interest to know what the $E_{initiation}$ values for continuous streamers would
be at the lower pressures found in the regions where lightning initiates. 
We therefore calculated $E_{initiation}$ for continuous streamers for a range of
pressures.  

The variation of $E_{initiation}$ for continuous streamers with both pressure
and drop size is shown in Fig \ref{E3d}.  The dark region in the lower left corner
indicates the region in which initiation is most favorable - large L and low
pressure. Over the chosen ranges of pressure and L, pressure has a greater
effect on $E_{initiation}$ than L.

Fig \ref{E3d} indicates that $E_{initiation}$ varies linearly with pressure. We
consider the dependence of the various parameters used by the model: $\alpha,
\eta, D$ and $v$ are functions of $E/p$ while the $\mu$ and $f_1 \cdot f_2$ are
functions of $l \cdot p$. The linear dependence of $E_{initiation}$ for 
continuous streamers suggests that the dependence on $E/p$ dominates and that
there is a unique value of the ``reduced'' E-field, $Y_{initiation}$ =
$E_{initiation}/p$  for a particular $E$ and $p$ combination.

\bhead{Propagation}
\label{prop}

\begin{figure}[ht]
\vspace{8mm}
\begin{center}
\includegraphics{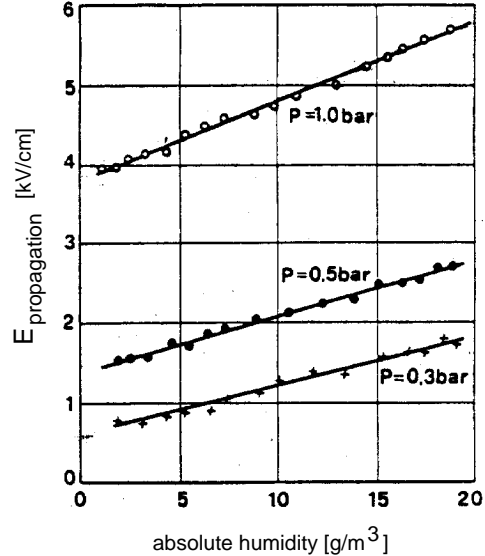}
\end{center}
\vspace{-8mm}
\caption{ $E_{propagation}$ as a function of absolute humidity and pressure,
$p$ \cite{griffithsB}.}
\label{Epropagation}
\end{figure}

The E-field necessary to sustain stable streamer propagation,
$E_{propagation}(p)$, was measured by \cite{griffithsB} as a function of air
pressure, $p$.  These stable streamers, once initiated, will continue to
propagate provided $E_{initiation} \ge E_{propagation}$. Griffiths and Phelps found that
$E_{propagation} \sim 400$ kV/m for dry air at $p=1000$ mb and that
$E_{propagation}(p) \propto p^5$ (Fig \ref{Epropagation}). At $p=500$ mb
$E_{propagation} \sim  150$ kV/m for dry air. 

At $p=1000mb$ Fig \ref{E3d} shows that $E_{initiation} > 400$ kV/m for all L. 
These initiated streamers will therefore be able to propagate over the entire length of the
region in which $E_{external}$ remains constant. In thunderclouds this scale is
typically hundreds of meters. At lower pressures $E_{initiation} >
E_{propagation}$ over a large range of L.  

\ahead{Discussion}

In this paper we have shown that continuous, propagating streamers can be initiated
from water drops at E-fields found in thunderstorms. Provided that
$E_{initiation}(p) \ge E_{propagation}(p)$, these streamers are capable of
propagating over considerable distances. This distance is limited by the size
of the region in which $E_{external}$ is greater than $E_{propagation}(p)$. 

When the electron currents in streamers become large enough, Joule heating
produces a 'warm' leader; a channel in which thermodynamic equilibrium is
destroyed and hydrodynamic effects become important.  This is commonly referred
to as the 'stepped-leader' in the cloud-to-ground lightning context. The currents carried by
individual streamers initiated at the drops are several orders of magnitude too low to
produce leaders \cite{anne}. These streamers may, however, still eventually lead
to leaders if they can be combined or multiplied. 

Griffiths \& Phelps [1976] considered the role of small scale discharges in
thunderclouds, calculating the E-field enhancement due to multiple propagations
of positive streamers near an electrode. According to their model, a series of three to
seven streamers gave rise to an enhanced E-field of up to $\sim 1500$ kV/m in a
region of several meters in linear scale near the electrode. It is possible that
several continuous streamers initiated from drops in the thundercloud could
provide the required field enhancement. \cite {griffiths} found that the field
was intensified on a time scale of $\sim$ 1 ms, which is comparable to the
lifetime of the coalesced drops as measured by CL. Further investigation is
required to determine whether a hydrometeor is capable of initiating multiple
streamers. 

An alternate mechanism for leader formation would be the combination of several
streamers in close proximity to one another to form a single, more vigorous
streamer with sufficient current to transform it to the ``warm'' leader stage.
If we think of drops that initiate continuous streamers as ``electrodes'' then
the number of ``electrodes'' available increases with increasing E (see Fig
\ref{EvsL}). Thus the likelihood of several streamers initiating in close
proximity increases and the chance of leader formation is increased. This is
also in keeping with the observations of large amounts of corona activity in
thunderstorms without lightning. Only if the ``electrode'' density is
sufficiently high will streamers be able to merge and form a leader. These
possible mechanisms for leader formation require further investigation.

The streamers observed by both CL and examined in our model were
all positive, occuring at the lower end of drops. This corresponds to drops located
above the negative charge center in clouds. Leader formation in this region is observed
to lead to intra-cloud lightning flashes. Drops located below
the negative charge center have negatively charged lower ends and
investigation of this situation will require the modeling of negative streamers
which are much more complex in nature than positive streamers
\cite{bondiou-ve}. Leader formation in this region (below the negative charge center)
will lead to cloud to ground lightning. No attempt has been made in this paper
to model these negative processes but future attempts should investigate this phenomena.

Finally, while we have concentrated on liquid hydrometeors, future work should
incorporate ice particles as possible ``electrodes''. This may explain how
lightning initiation can occur at higher altitudes near the upper positive
charge center in thunderclouds where there is little or no liquid water available. 
\vspace{1cm}

\acknowledgements
We are grateful for support by NASA \# NAG8-1150. We thank Anne
Bondiou-Clergerie of ONERA for supplying the ionization and attachment coefficient data
and providing helpful comments and advice. We are also grateful to Ron Geballe
for his suggestions.

\end{document}